\documentclass[10pt,letterpaper]{article}
\usepackage[top=0.85in,left=2.75in,footskip=0.75in]{geometry}
\usepackage[utf8]{inputenc}
\usepackage[english]{babel}
\usepackage{csquotes}
\usepackage{amsmath,amssymb}

\usepackage{textcomp,marvosym}

\usepackage{nameref,hyperref}

\usepackage[right]{lineno}

\usepackage[table]{xcolor}

\usepackage{array}
\usepackage{siunitx}

\newcolumntype{+}{!{\vrule width 2pt}}

\newlength\savedwidth

\usepackage{setspace} 
\raggedright
\usepackage[aboveskip=2ex,labelfont=bf,labelsep=period,justification=raggedright,singlelinecheck=off,belowskip=1ex]{caption}

\usepackage[%
    backend=biber,
    style=nature,
    url=false,
    doi=false,
    date=year,
    isbn=false,
    arxiv=false,
    eprint=false,
]{biblatex}
\usepackage{lastpage,fancyhdr,graphicx}
\usepackage{epstopdf}
\fancyheadoffset[L]{2.25in}
\fancyfootoffset[L]{2.25in}
\begin{document}
\vspace*{0.2in}

\begin{flushleft}
{\Large
\textbf\newline{Inferring the effect of interventions on COVID-19 transmission networks} % Please use "sentence case" for title and headings (capitalize only the first word in a title (or heading), the first word in a subtitle (or subheading), and any proper nouns).
}
\newline
\\
Simon Syga\textsuperscript{1},
Diana David-Rus\textsuperscript{2},
Yannik Schälte\textsuperscript{3,4},
Michael Meyer-Hermann\textsuperscript{5,6},
Haralampos Hatzikirou\textsuperscript{1,7},
Andreas Deutsch\textsuperscript{1,*}
\\
\bigskip
\textbf{1} Center for Information Services and High Performance Computing,
Technische Universität Dresden, Nöthnitzer Straße 46, 01062 Dresden, Germany
\\
\textbf{2} Bavarian Health and Food Safety State Authority (LGL) -- Research division, 
 Veterinärstraße~2, 85764, Oberschleißheim, Germany
\\
\textbf{3} Institute of Computational Biology, Helmholtz Zentrum München –
German Research Center for Environmental Health, 85764 Neuherberg, Germany
\\
\textbf{4} Center for Mathematics, Technische Universität München, 85748 Garching, Germany
\\
\textbf{5} Department of Systems Immunology and Braunschweig Integrated Centre of Systems Biology (BRICS), Helmholtz Centre for Infection Research, Braunschweig, Germany
\\
\textbf{6} Institute for Biochemistry, Biotechnology and Bioinformatics,
Technische Universität Braunschweig, Braunschweig, Germany
\\
\textbf{7} Mathematics Department, Khalifa University, P.O. Box 127788, Abu Dhabi, United Arab Emirates
\\
\bigskip
*andreas.deutsch@tu-dresden.de
\end{flushleft}
\section*{Abstract}
Countries around the world implement nonpharmaceutical interventions (NPIs) to mitigate the spread of COVID-19. 
Design of efficient NPIs requires identification of the structure of the disease transmission network. 
We here identify the key parameters of the COVID-19 transmission network for time periods before, during, and after the application of strict NPIs for the first wave of COVID-19 infections in Germany combining Bayesian parameter inference with an agent-based epidemiological model. 
We assume a Watts-Strogatz small-world network which allows to distinguish contacts within clustered cliques and unclustered, random contacts in the population, which have been shown to be crucial in sustaining the epidemic.
In contrast to other works, which use coarse-grained network structures from anonymized data, like cell phone data, we consider the contacts of individual agents explicitly.
We show that NPIs drastically reduced random contacts in the transmission network, increased network clustering, and resulted in a change from an exponential to a constant regime of new cases. 
In this regime, the disease spreads like a wave with a finite wave speed that depends on the number of contacts in a nonlinear fashion, which we can predict by mean field theory. 
Our analysis indicates that besides the well-known transition between exponential increase and exponential decrease in the number of new cases, NPIs can induce a transition to another, previously unappreciated regime of constant new cases.% line 142 and 145 can be compressed in one line. Basically they are saying same thing. 

\section*{Introduction}%\the\textwidth
\begin{figure}[!ht]
\centering
\includegraphics[width=.9\textwidth]{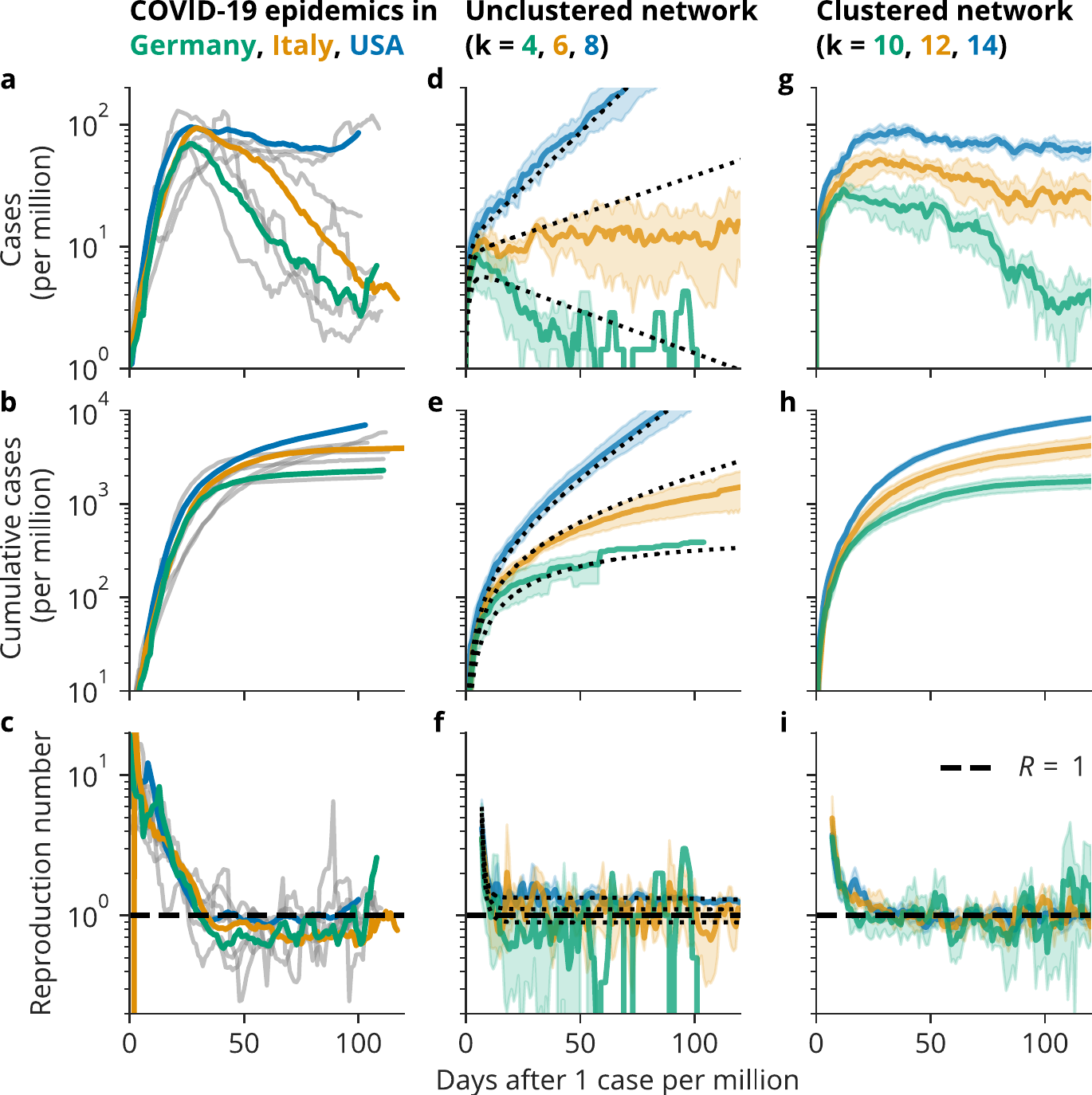}
    \caption{\textbf{Dynamics of the COVID-19 epidemics can be explained by highly clustered transmission networks}. (\textbf{a-c}) New cases per million, cumulative cases per million and reproduction number in Germany, Italy and the US. Gray lines are other countries for comparison. After the implementation of NPIs, the case numbers decrease slowly, corresponding to effective reproduction numbers of just below 1. Own visualization of data from Johns Hopkins University \cite{Dong2020}. (\textbf{d-f}) Disease dynamics in a random network. The number of new cases and cumulative cases changes exponentially over time, strongly depending on the number of contacts and the infection probability. The reproduction number is equal to one only for a fine-tuned set of parameters. Black dotted lines correspond to predictions of the respective ODE approximation. (\textbf{g-i}) Disease dynamics in a strongly clustered small-world network ($p \approx 0$). After an initial exponential increase in cases, the number of new cases is almost constant over time, corresponding to a linear increase in cumulative cases and reproduction numbers around 1. This behavior is robust against changes in the total number of contacts $k$ and the infection probability $p_I$, see text.  (\textbf{d-i}) show mean and standard deviation of 5 independent simulations per parameter set on Watts-Strogatz networks with $n = 10^5$ nodes.\label{fig:data}}
\end{figure}

The SARS-CoV-2 pandemic has dramatic consequences at a global scale. 
Until herd immunity has been reached through vaccination, countries rely on non-pharmaceutical interventions (NPIs) of varying severity, like canceling big events, closing schools, and shutting down businesses to reduce virus transmission. 
An important goal of NPI design is to prevent those contacts in the population which contribute the most to disease spread while allowing less dangerous contacts. 
Data from several countries indicate that the effect of early, less strict NPIs, for example cancellation of large events, had a profound effect on the spread of the disease \cite{brauner2021}. 
Other findings suggest that only a full lockdown reduced the spread noticeably \cite{Flaxman2020}, although the implementation of the lockdown varied dramatically between individual countries. 
For example, during the first wave, Germany did not implement a full lockdown but enacted contact restrictions and closure of nonessential businesses so that people were still allowed to leave their homes and meet in small groups.
In several countries, following the implementations of various NPIs, the curves of cumulative infections left the exponential regime and entered a linear one, corresponding to a constant regime of new infections, and an effective reproduction number around 1, see Fig.~\ref{fig:data}\textbf{a-c}.
This is remarkable, because these countries are quite heterogeneous regarding their demographics, economic situation and the implemented NPIs.
Epidemiological models that assume an underlying random transmission network predict this behavior only for a particularly fine-tuned set of parameters, which contrasts with the robustness of the decline in the reproduction number observed in reality \cite{Kermack1927,Ridenhour2014,NeilMFergusonetal.2020,Dehning2020}.
Crucially, this is still true for detailed compartment models that incorporate the effect of test-trace-isolate (TTI) efforts, asmptomatic spreaders, or age-dependent spreading based on contact matrices \cite{aleta2020,khailaie2021} and for agent-based models that rely on coarse-grained contact networks, for example created from anonymized cell phone data, which assume a well-mixed situation on a mesoscopic scale of hundreds of agents \cite{rader2020,lau2020}.
Thurner et al. suggested that the linear regime of cumulative infections is a consequence of small-world transmission networks with high clustering \cite{Thurner2020b}, see Fig.~\ref{fig:data}\textbf{d-i}.
A similar model was able to explain the disease dynamics during the SARS outbreak in Hong Kong in 2003 \cite{Small2005}.
Komarova et.~al. discussed the power law behavior of the dynamics of COVID-19 spread in the context of a metapopulation model \cite{Komarova2020}, while the power law dynamics during the hard lockdown in Chinese provinces could be explained by a compartment model that assumes that the susceptible population was quarantined on a time scale comparable to the infectious period of the disease, so that the epidemic comes to a halt quickly \cite{maier2020}.

Identification of the underlying transmission network is not only crucial for the design of effective NPIs, but also for assessing other properties of COVID-19 spread, such as the herd immunity threshold \cite{Britton2020}.
The heterogeneous topology of real social networks is reflected by a small average path length between any two nodes (small world property), a high clustering in the network (the probability of two nodes being connected is much larger if they have a neighbor in common) and by a power-law distribution (scale-free property) of the node degree \cite{Watts1998,Barabasi2003a}.
Network-based epidemiological models allow to consider the effects of heterogeneity with respect to the type and frequency of contacts in the population, i.~e. how often people meet and whom, by representing all agents as nodes of a network and the contacts in the population by links between these nodes \cite{Keeling2011,Pastor-Satorras2015}. 
For example, the spread of diseases is strengthened on scale-free networks so that the epidemic threshold is reduced \cite{Pastor-Satorras2001,Pastor-Satorras2002,Moreno2002}.

Here, we argue that during the period of severe NPIs, like contact restrictions, the most important feature of real transmission networks is their strong clustering.
This means that because public places and events are closed, we expect that people focus their contacts on a single group (clique), where almost each member of this clique is contact with each other.
Typically examples for such cliques include households or colleagues at work.
On the other hand, we will neglect the scale-free property of social networks, because it requires that there are a few people with a very large number of contacts, for example at events, schools, large private gatherings etc., which are the targets of most NPIs.
We combine Bayesian parameter inference \cite{Minter2019} with an epidemiological model based on the Watts-Strogatz small-world network \cite{Watts1998} that allows to interpolate between unclustered and highly clustered transmission networks, see Fig.~\ref{fig:network_sketch}, to infer the topology of the transmission network in Germany during three time periods: February 26 until March 15, before serious NPIs were imposed, March 16 until June 6, when strict contact restrictions were in place and nonessential businesses were closed, and June 7 until September 15, when most NPIs were lifted.
We show that NPIs reduced random contacts in the transmission network, resulting in a change from an exponential to a constant regime of new cases.
Random contacts often span a large distance in the transmission network and connect different cliques.
They include, for example, contacts in public transport, bars and restaurants, but also contacts with relatives that live far away.
Furthermore, given the nature of random contacts, the probability for superspreading events increases when there is a high density of such random contacts, as they enable the disease to spread to fully susceptible cliques.
Moreover, we show that the reduction of random contacts has a strong nonlinear effect on the effective reproduction rate and the peak value of infected people, pushing the disease dynamics into a regime of a constant number of new infections, see Fig.~\ref{fig:data}\textbf{g-i}.
Decreasing the total number of contacts including those within connected cliques, on the other hand, has a strong effect if the epidemic threshold is reached or if the network has a large fraction of random contacts and weak clustering, see Fig~\ref{fig:data}\textbf{d-f}.
Furthermore, we are able to analytically predict disease spread in the regime of almost no random contacts by a wave equation where the wave speed (proportional to the growth rate) $c$ scales as $c \propto k \sqrt{k}$, where $k$ is the number of clustered, local contacts.
This corresponds to the linear spreading regime.
In summary, our model suggests that NPIs decrease random contacts in the population, can prevent an exponential disease spread and induce a previously unappreciated transition to a regime of constant new cases.%DDR--ok I don't want to be %'devil's advocate' here, but this in principle is not also what the paper of Thurner et. al concluded as you already %mentioned in line 167? I would  here also add the bayesian inference method that we use that you mention in line 199

\section*{Results}
\subsection*{Bayesian parameter inference} 
\begin{figure}[!ht]
\centering
\includegraphics[width=.9\textwidth]{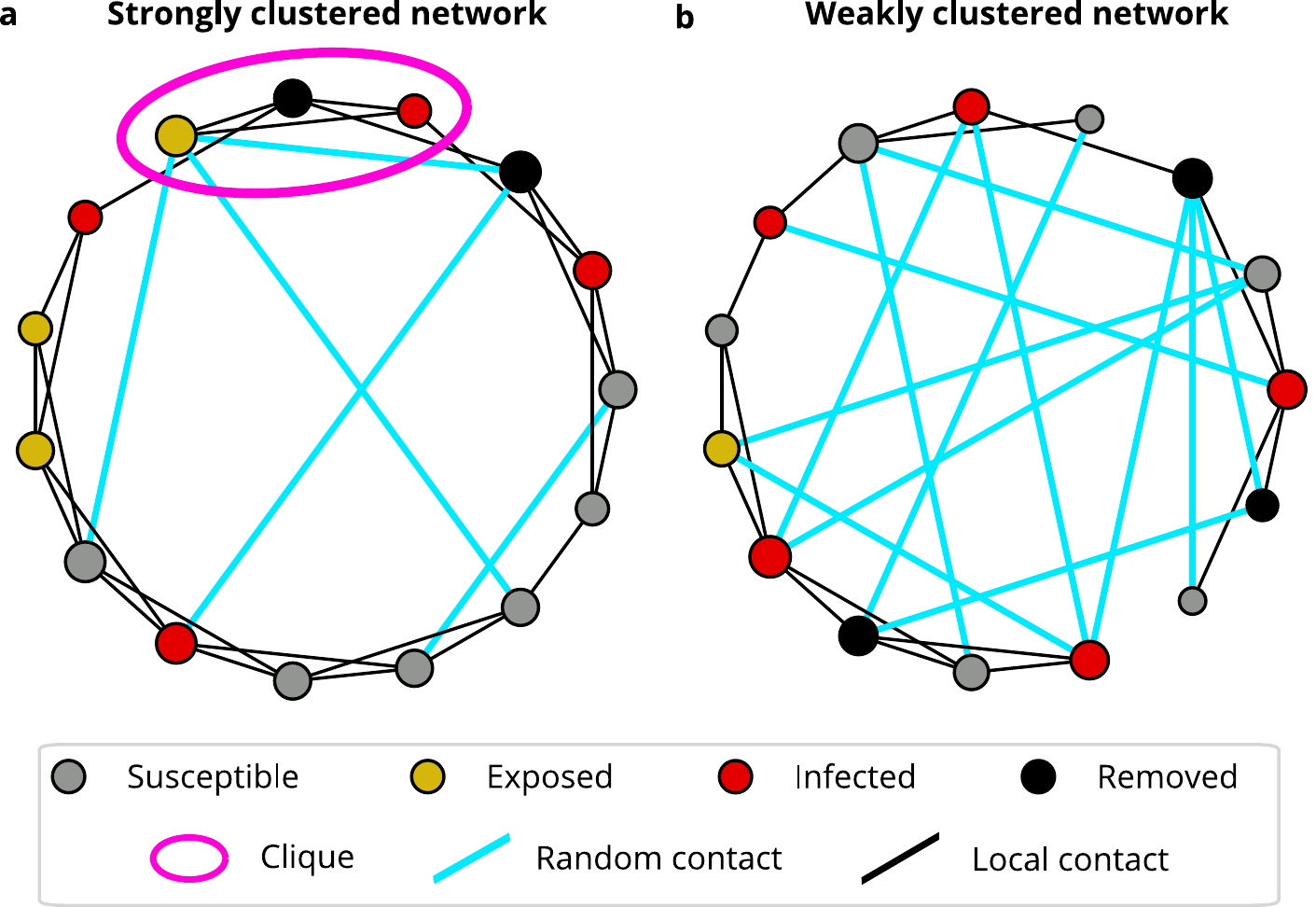}
   \caption{\textbf{Watts-Strogatz small world network.} Agents are placed in a ring-like topology and linked to their $k$ nearest neighbors (black lines). Next, every link is rewired randomly with a small probability $p$ (cyan lines). Every agent has one of four states: susceptible (gray), exposed (gold), infectious (red), or removed (black). Infectious agents spread the disease to connected susceptible agents with a probability $p_I$ in each time step. The size of network nodes is proportional to the node degree. (\textbf{a}) Strongly clustered network. Almost all contacts are restricted to neighbors ($p = 0.1$). (\textbf{b}) Weakly clustered network with a large fraction of random contacts ($p=0.5$). Other parameters are constant are equal in (\textbf{a}) and (\textbf{b}): $n = 15$ agents, $k = 4$ contacts. \label{fig:network_sketch}}
\end{figure}

We aimed to infer the induced changes in the topology of the COVID-19 transmission network by Bayesian parameter inference. 
We expected that NPIs lead to a change in the behavior of people and therefore in the topology of the corresponding transmission network.
We assumed that the transmission network can be described by the Watts-Strogatz network \cite{Watts1998} that can interpolate between a weakly and a strongly clustered small-world network, see Fig.~\ref{fig:network_sketch}. 
Crucially, in this framework, we could distinguish local, clustered contacts within cliques, like households, nursing homes, businesses, etc. and random contacts outside of these clusters, corresponding to encounters in public transport, with business partners, friends and family that live far away and similar. 
During the construction of the Watts-Strogatz network, $n$ nodes are placed in a ring topology and connected to their $k=2,4,6,\dots$ nearest neighbors (local contacts, black lines in Fig.~\ref{fig:network_sketch}).
After that, each link is rewired with a probability $p$ to another random node (random contacts, cyan lines in Fig.~\ref{fig:network_sketch}).
We used the SEIR (susceptible-exposed-infectious-removed) epidemiological model for the COVID-19 disease dynamics.
Agents in the SEIR model were represented by network nodes such that infectious agents could spread the disease with probability $p_I$ to the susceptible agents that they are connected to in discrete time steps of single days (Methods, Fig.~\ref{fig:network_sketch}, Supplementary Fig.~1).

We inferred the model parameters $p, k, p_I$ for the time periods before and after the first NPIs were implemented in Germany, and after most NPIs were lifted again using an approximate Bayesian computation with sequential Monte Carlo (ABC-SMC) algorithm (Methods).
We did not infer those SEIR model parameters corresponding to the disease progression, because they were reported in the literature as \SI{3+-1}{d} to become infectious after exposure and \SI{10+-3}{d} of infectiousness \cite{Linton2020} and kept the number of agents fixed at $n = 3  \times 10^5$, which we regard as a representative sample of the whole population.
Note that we do not explicitly account for the quarantine of infectious agents, which could be done by a time period-dependent removal probability, for example.
However, we do not do this, because the infection and removal probabilities cannot be determined independently \cite{Dehning2020} and there is no reliable data on the effectiveness of TTI measures to further specify the removal probability.

For the time periods before June 6, for which a large number of infections was undetected, we also inferred the initial numbers of exposed and infectious individuals $n_E(0), n_I(0)$.

Based on Google mobility reports and previous work on the inference of change points in the spread of COVID-19 \cite{Google,Dehning2020} we assumed the critical time point for the effect of NPIs in Germany to be March 15, as from March 16, NPIs were synchronized in German states, and schools and nonessential businesses were closed.

\begin{figure}[ht!]
\centering
\includegraphics[width=0.9\textwidth]{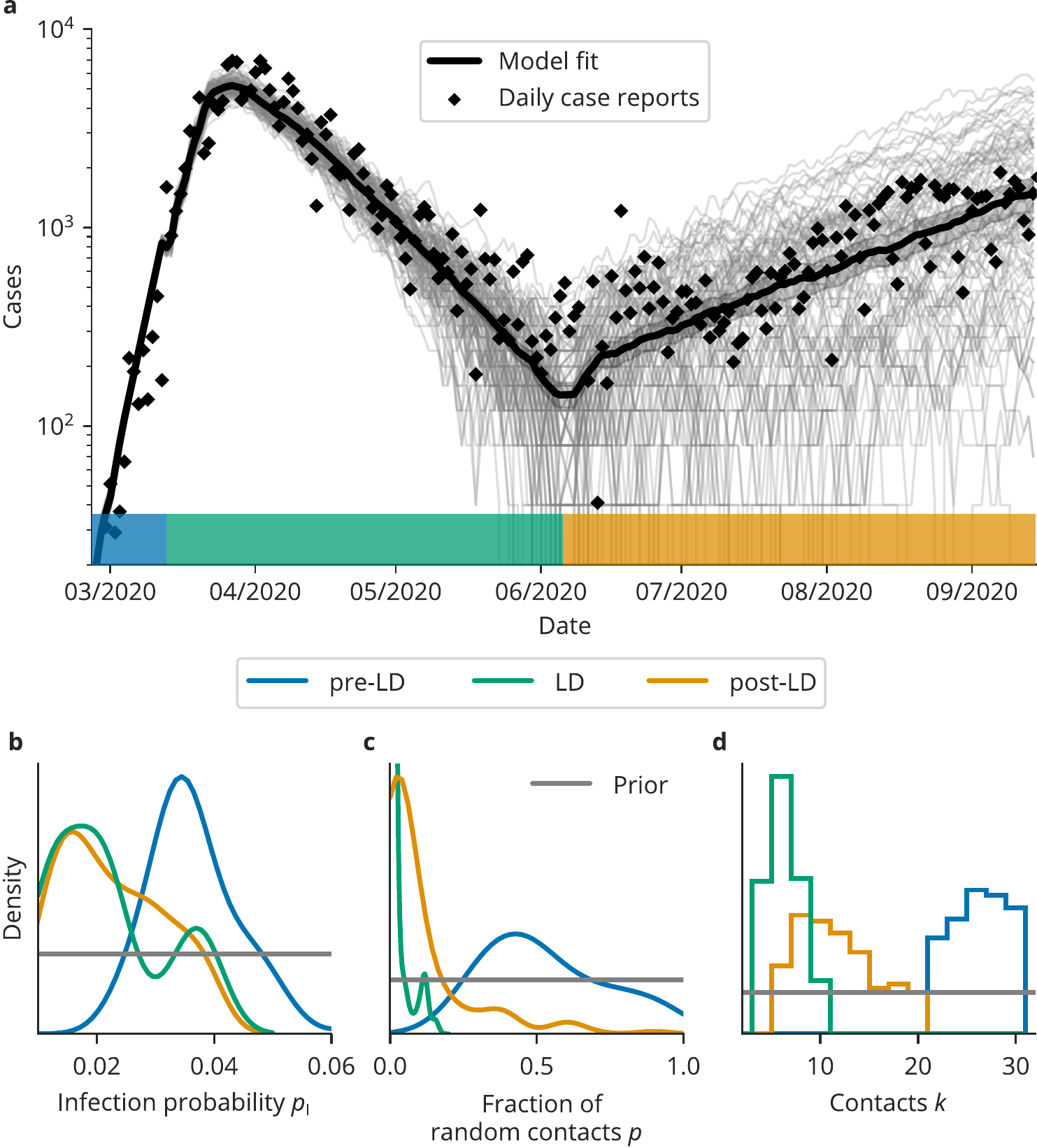}
    \caption{\textbf{Inference of key epidemiological parameters before (pre-LD, blue), during (LD, green) and after (post-LD, yellow) the lockdown in Germany}. (\textbf{a}) Daily case reports (black diamonds), single model instances (gray lines) and their mean (black line, error band corresponds to 95 \% error of the mean). Colored bars indicate the respective time periods, where the color is the same as that of the corresponding posterior parameter distributions in (\textbf{b-d}). Model parameters were chosen as the median of the respective posterior distribution. (\textbf{b} - \textbf{d}) Kernel density estimates of the posterior parameter distributions for the three time periods. The uniform prior distribution is shown in gray. (\textbf{b}) The infection probability $p_I$ decreased from $p_I = 0.035$ (CI $[0.025, 0.061]$) pre-lockdown to $p_I = 0.02$ (CI $[0.01, 0.04]$) during the lockdown. After most restrictions were lifted, the infection probability remained almost unchanged at $p_I = 0.02$ (CI $[0.01, 0.04]$). (\textbf{c}) The fraction of random contacts in the transmission network $p$ decreased strongly from $p = 0.48$ (CI $[0.23, 0.94]$) to $p = 8 \times 10^{-5}$ (CI $[10^{-7}, 0.12]$) when restrictions were put in place and increased to $p = 0.03$ (CI $[0.001, 0.60]$) when they were lifted. (\textbf{d}) The total number of contacts $k$ decreased strongly from $k = 26$ (CI $[22, 32]$) to $k = 6$ (CI $[4, 10]$) during lockdown before increasing again to $k = 12$ (CI $[6, 20]$) after restrictions were lifted.
    \label{fig:abc}}
\end{figure}

We intentionally chose broad, uninformative priors for all parameters, such that we could compare the obtained posterior distributions with other data sources as a sanity-check of our approach.
To account for the weekday-dependent reporting delay, we used a seven day rolling average of new case reports provided by Johns Hopkins University \cite{Dong2020}, see Fig~\ref{fig:abc}\textbf{a}.
Our parameter inference scheme is based on a minimization of the difference between this average and the number of agents becoming infectious in the model on the corresponding day. 
For the initial phase, we base our inference on the absolute number of infections, while we used the relative number for the following time periods (Supplementary Information).

We here assumed that all people that were tested positive were also infectious at the time of the test and that the time delay between people turning infectious and their test is negligible.

First, we inferred the parameters for the time from February 26 to March 15, since daily new cases increased rapidly after February 26, while there were almost no cases in the week before.
Our analysis of this period revealed an almost random transmission network, with a median fraction of random contacts of $p = 0.48$ (with 95~\% credibility interval, CI $[0.23, 0.94]$), a large number of contacts $k = 26$ (CI $[22, 32]$) and a high infection probability of $p_I = 0.035$ (CI $[0.025, 0.061]$).
The inference of this unexpectedly high number of contacts could be the result of a scale-free degree distribution before NPIs were imposed (Discussion).
For the initial condition we estimated that $33$ (CI $[4, 46]$) people were exposed and $81$ (CI $[32, 118]$) people were infectious on February 26.

To assess the changes of the transmission network induced by the NPIs in Germany, we next considered the time period following March 16. 
During that time frame, the NPIs were changed several times, however, the contact restrictions, which we regard as the most crucial intervention, were only lifted on June 6, which is why we chose this date as the endpoint of the time interval.
As the total number of cases was computationally intractable, we here used the relative number of new cases as input for this time period (Supplementary Information).

Our Bayesian parameter inference reveals that the NPIs reduced the number of contacts in the transmission network considerably to $k = 6$ (CI $[4, 10]$).
They also reduced the infection probability of these contacts to $p_I = 0.02$ (CI $[0.010, 0.039]$), which matches well with an estimation based on the individual-level secondary attack rate in the household of 17~\% \cite{Jing2020}.
Crucially, the fraction of random contacts decreased to $p = 7\times 10^{-5}$ (CI $[10^{-7}, 0.12]$), stopping the exponential growth. 
Additionally, we estimated the number of exposed people on March 16 to be $n_E(\mathrm{March 16}) = 78$ (CI $[31, 147]$)  per million and the number of infectious people to be $n_I(\mathrm{March 16}) = 452$ (CI $[301, 656]$) per million.  
The fact that during the week before March 16 there were only 57 infections per million detected in Germany is a hint that a large fraction of infections went unnoticed at the time, which agrees with other reports \cite{Streeck2020,chen2021}. 

We also inferred the model parameters for the time period following June 6 when contact restrictions were lifted.
To this end, we used the final time point of simulation instances from the previous period as initial conditions.
For the number of contacts we obtained a median $k = 12$ (CI $[6, 20]$) that matches well with reports of the average number of daily contacts in Europe of 13.4 \cite{Mossong2008}.
The median fraction of random contacts was estimated as $p = 0.03$ (CI $[0.001, 0.6]$), which means it was notably smaller than before NPIs had been implemented, but larger than in the time period of strict NPIs.
Interestingly, we found that the infection probability was as low in this time period at $p_I = 0.02$ (CI $[0.01, 0.04]$) as during the lockdown, which could be the result of a seasonal effect and of people spending more time outside, which hinders the spread of airborne diseases such as COVID-19.
We inferred rather broad parameter posterior distributions for this period, due to the generally low number of infections and large localized outbreaks, leading to a large variation in daily case counts.
This is also reflected in a big variability between single model instances for this time period, see Fig.~\ref{fig:abc}\textbf{a} (gray lines).
For the time periods without strict NPIs (February 26 to March 15 and June 6 to September 15), for which we inferred a large fraction of random contacts, the infection probability $p_I$ and the number of contacts $k$ were notably correlated, leading to broader posterior distributions. 
Figure~\ref{fig:abc} shows the model fit compared to the daily case counts and the corresponding prior and posterior parameter distributions of the network parameters and the infection probability for the three time periods.
\subsection*{Disease dynamics in the clustered network}
\begin{figure}[ht!]
\centering
\includegraphics[width=.9\textwidth]{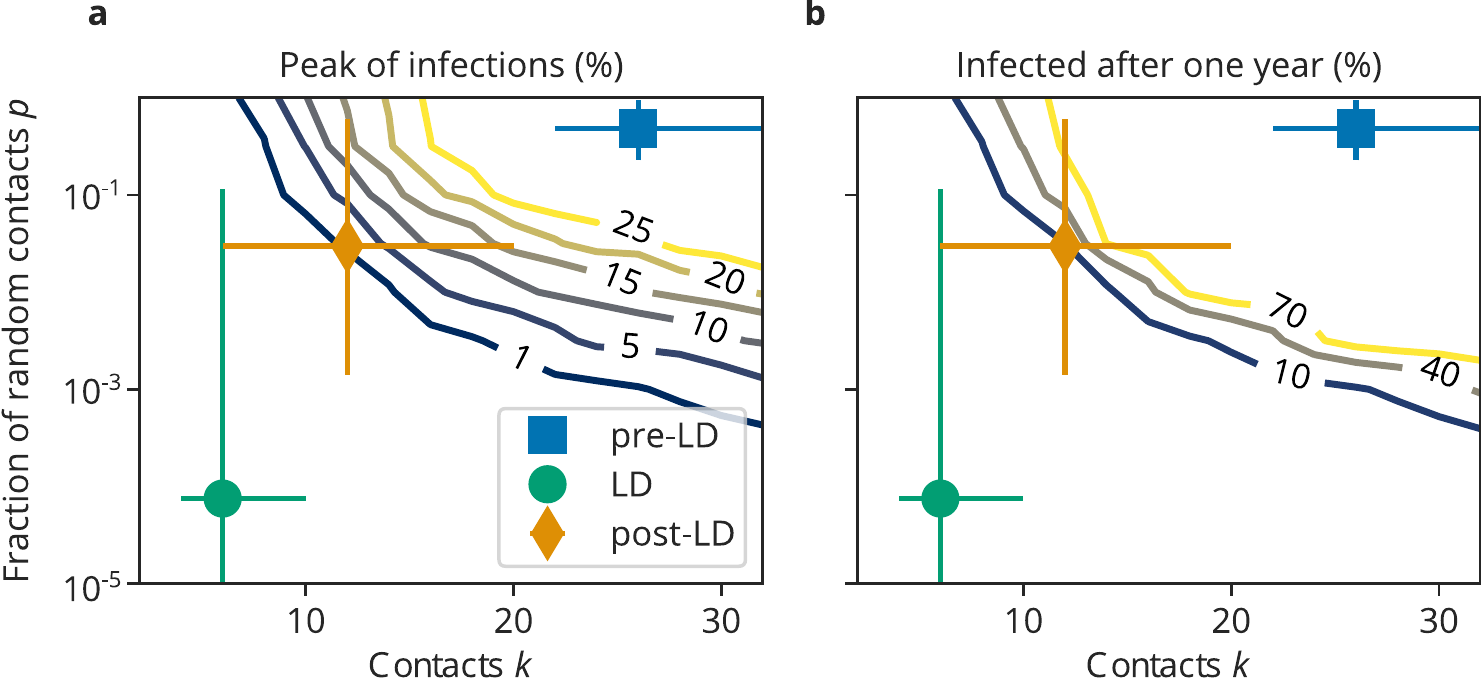}
    \caption{\textbf{NPIs mitigated the disease spread by reducing random contacts.} (\textbf{a}) Number of simultaneously infected people (peak of infections) in percent of population. The wave peak can be massively mitigated by decreasing the fraction of random contacts, even while keeping the total number of contacts constant. (\textbf{b}) Cumulative number of infections after one year in percent of the population. Similar to the peak of infections, the cumulative number of infections can be limited by reducing the fraction of random contacts. Blue square (pre-LD), green point (LD) and yellow diamond (post-LD) correspond to the median parameters and 95~\% CI obtained from Bayesian parameter inference for the time periods 26/02--15/03, 16/03--05/06 and 06/06--15/09, respectively and are shown as a reference. The NPIs after March 15 prevented an exponential spread of the disease, but lifting them led to another exponential increase. Shown are contour lines of the mean of 20 independent model realizations of each parameter combination $(p,k)$, while the other parameters were fixed at $n=10^5, p_I = 0.02$.}
    \label{fig:quantifications}
\end{figure}

To determine the transitions between the linear and exponential regimes of the disease dynamics, we performed a parameter scan varying the network parameters $p$ and $k$, while keeping the disease-specific parameters and the system size fixed at $n = 10^5, p_I = 0.02$.
Thereby, reducing $k$ corresponds to reducing the total number of contacts (local and random), while reducing $p$ does not change the number of contacts, but restricts them to locally clustered agents (cliques).
We recorded the peak number of simultaneously infected people, because a central goal of NPIs is to prevent the overload of the health system, and the total number of infected people after 100 days as a measure for the total damage to public health, see Fig~\ref{fig:quantifications}.

Importantly, the number of infections could be reduced massively by only decreasing the fraction of random contacts in the population, while keeping the total number of contacts constant.
As we have inferred the parameters of the transmission network for different time periods, we could associate them with regions in our parameter space.
If no NPIs had been implemented and people would not have changed their behavior, more than 40~\% of people could have been infected simultaneously and almost everybody would have been infected after one year (Fig~\ref{fig:quantifications}, blue square).
The peak of infections was reduced to 0.0002~\% by the interventions (Fig~\ref{fig:quantifications}\textbf{a}, green point).
Lifting the NPIs moved the system back into the exponential regime, with a projected peak of infections of 1.3~\% and almost 10~\% of the population to be infected within one year (Fig~\ref{fig:quantifications}, yellow diamond).
Reducing the total number of contacts can in principle push the system below the epidemiological threshold leading to extinction of the disease (see Figs.~\ref{fig:data}, \ref{fig:quantifications}), however the effect is weaker in the regime far away from the threshold (see Figs.~\ref{fig:data},\ref{fig:quantifications}, $k = 10, 12, 14$).
On the other hand, in the strongly clustered regime $p \approx 0$, increasing the fraction of random contacts has a dramatic effect: both the peak value of infected agents and the total number of infected agents increase in a non-linear manner.
Preventing most random contacts in the network ($p\to0$) hinders the spread of the disease, so that the effective reproduction number fluctuates around 1, and the cumulative number of infections increases linearly with time as observed in many countries after the first NPIs were imposed.
\subsection*{Wave speed of infections in the linear regime}
\begin{figure}[ht!]
\centering
\includegraphics[width=.9\textwidth]{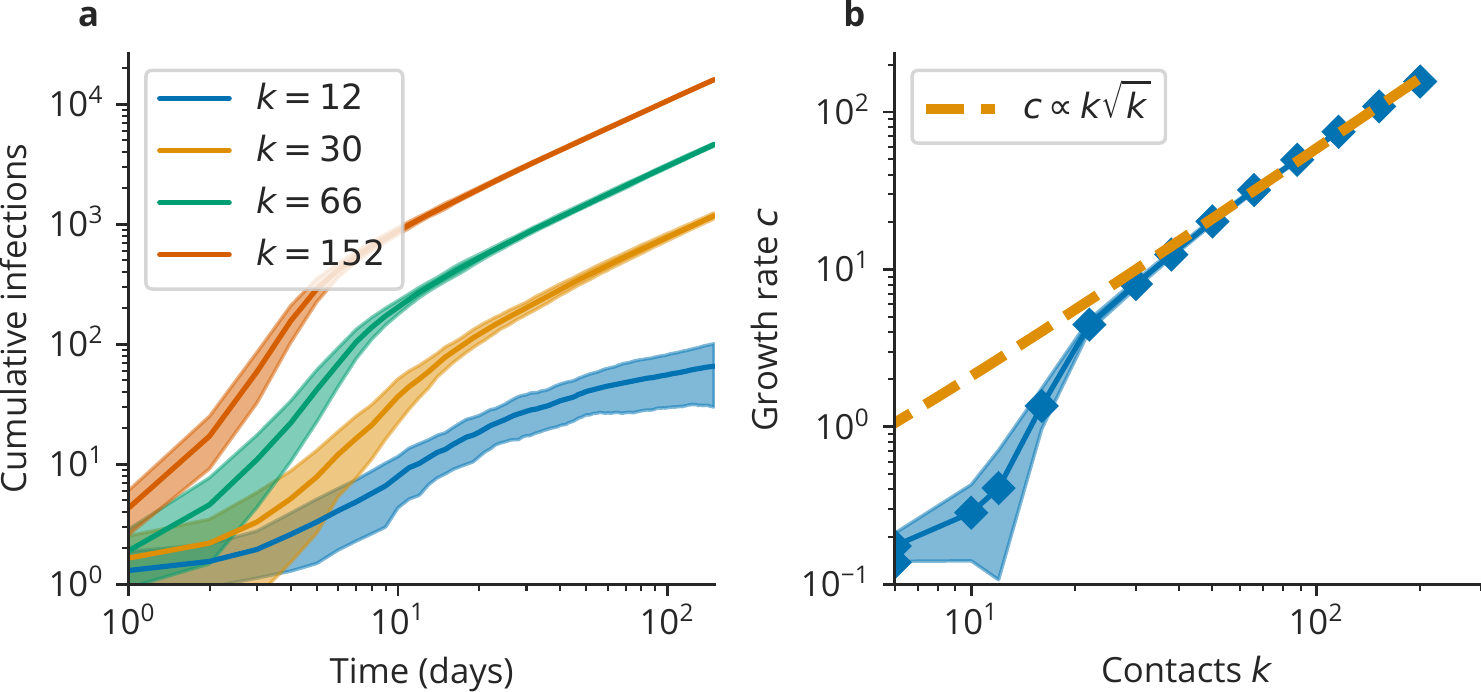}
    \caption{\textbf{The growth rate of the cumulative number of infections in the linear regime can be predicted analytically}. (\textbf{a}) Cumulative number of infections in the linear regime ($p=0$) in the network-based model. (\textbf{b}) Growth rate of cumulative infections in dependence of the number of contacts $k$. The growth rate scales as $c\propto k \sqrt{k}$ as predicted by Eq~\ref{eq:cmin} for a large number of contacts $k$. Shown is the mean and standard error of the mean (SEM) of 20 independent simulations for each parameter.}
    \label{fig:wavespeed}
\end{figure}

We were especially interested in the disease dynamics in the regime $p \to 0$, as this is where traditional epidemiological models that assume a random transmission network break down.
In the case of only local contacts in the network, the disease spreads like a wave originating from the initially infectious agent.
This wave-like disease spread was also reported in real networks, such as the air traffic network \cite{Brockmann2013}.
To calculate the speed of the infection wave, we used a mean-field approximation of an SIR-like agent-based model operating on the Watts-Strogatz network (Supplementary Information). 
We scaled the mean-field equations to continuous time $t$ and space $x$ (where the distance $\Delta x$ is measured as the number of links along the ring of nodes), and approximated the dynamics by a set of partial differential equations for the probability densities of susceptible $\sigma(x,t)$, infectious $\iota(x,t)$ and removed $\rho (x,t)$ agents (Supplementary Information).
In particular, we obtained the following equation for the probability density of infectious agents
\begin{multline}\label{eq:iota}
    \partial_{tt} \iota (x,t) = \frac{2}{\tau} \left\{ -\partial_t \iota (x, t) +\kappa_I k \sigma (x,t) \left[ (1-p) \left( \iota(x,t)  \right) + p I(t) \right] - \kappa_R \iota(x,t) \right\} + \\ +
    \kappa_I k (1-p) \sigma (x,t) D_k \partial_{xx} \iota(x,t),
\end{multline}
where $\kappa_I$ is the infection rate, and $\kappa_R$ is the removal rate. 
Here, $\tau$ is the short time scale of the local disease dynamics of a single agent, while $D_k := \frac{\Delta x^2 \tilde{k}}{\tau}$, with $\tilde{k} := (k/2 + 1)(k + 1) / 12$ is a constant that determines the disease spread in the network on longer time scales. 
The total number of infectious agents $I$ is defined as $I := \int \iota \, \mathrm{d}x$, where the integral represents a nonlocal coupling by random contacts.
The equation resembles the Telegrapher's equation but with a nonlocal coupling by random contacts and a nonlinear diffusion term due to local contacts.
We recover the classical SIR model for $p = 1$, as expected.
For the regime $p=0$, we obtained the speed of disease spread through the network as
\begin{equation}\label{eq:cmin}
    c = \sqrt{\kappa_I k D_k} \propto k\sqrt{k}.
\end{equation}
Notably, the wave speed depends on $k$ in a nonlinear manner.
Comparing our prediction against simulation data (Supplementary Information) revealed that the wave speed is proportional to the growth rate of the cumulative infections in the linear regime, see Fig~\ref{fig:wavespeed}.
\section*{Conclusion}
% Summary
We used Bayesian parameter inference to quantify the effects of government interventions in Germany on the transmission network of COVID-19 assuming it can be approximated by a Watts-Strogatz network.
This network captures the key feature of social networks affected by NPIs, namely the strong clustering of contacts when people restrict their social life to a small, interconnected clique.
Our analysis revealed that NPIs lead to a reduction in transmission probability, number of contacts, and, crucially, to the removal of almost all contacts outside highly clustered cliques.
In contrast to standard epidemiological models, in this regime the cumulative number of infections does not increase exponentially but linearly, with a massively reduced peak of infections.
The dynamics corresponds to a wave-like spread of the disease in the network, whose wave speed $c$ we predicted by mean-field theory to scale as $c \propto k \sqrt{k}$ in dependence of the number of contacts $k$.
At the same time, the effective reproduction number fluctuates around 1, irrespective of the wave speed.
However, as long as the epidemic threshold is not reached by the reduction of contacts between cliques and the reduction of the infection probability, the disease still spreads in the population, which emphasizes the need for an effective test, trace, and isolate (TTI) system and, ultimately, a vaccine.

A similar system was studied in a theoretical work on epidemics with two levels of mixing \cite{Ball1997}.
There, the authors found that local transmission leads to an increase in the effective reproduction number proportional to the local outbreak.
This in turn means that a very small fraction of random connections in the population are enough to enable a global outbreak, which corresponds to the exponential regime in our work.
Several other studies have highlighted the importance of random contacts between members of different cliques.
A recent study which used mobility network data to show that the spread of the disease is mostly driven by infections at events, which connect different communities, for example, in restaurants and religious establishments, further supports the importance of random contacts \cite{Chang2020}.
Moreover, empirical studies of the circumstances under which people got infected revealed that although 46 to 66~\% of transmission is household-based (clustered contacts), random contacts between these cliques are essential to sustain the epidemic, even if only a low percentage of infections are caused directly by them \cite{Lee2020a}.
A theoretical study that investigated the effect of different social network-based distancing strategies showed that the most effective social distancing strategy is to restrict contacts to a single clique, and to eliminate any contacts between the cliques (random contacts) \cite{Block2020}.

We based our parameter inference on the daily count of positive tests. 
Clearly, this approach is not perfect: the number of positive tests depends on the testing policy and the number of available tests, and the date, when the test result is recorded is always delayed. 
Another option is the usage of daily deaths, which are independent of testing.
However, it is very difficult to base our approach on the death count, simply because the total death count in Germany during the first wave and especially during the summer was comparatively quite low. 
That means that even our scaled-down system needs to be intractably large, if we want to reproduce the number of deaths using realistic assumptions about the fatality rate of the disease.
Additionally, the time distribution between infection and death has a big variance of up to 10 days, as it varies depending on the age of the affected individual \cite{Linton2020,faes2020}, further increasing the noise in the already low death numbers.

In our model we assumed a Watts-Strogatz transmission network and distinguished between random and clustered links.
Clearly it is possible to use other algorithms to construct clustered networks, for example the stochastic block model \cite{holland1983}, the relaxed caveman graph \cite{fortunato2010}, or the configuration model with defined clustering \cite{Newman2009}, to name a few.
We chose the Watts-Strogatz graph for its simplicity and its low number of parameters, but we expect that one can obtain similar results using other graph ensembles, as long as they allow for high clustering and overlap of connected cliques. 

Moreover, we did not account for the scale-free degree distribution found in real social networks.
It is a result of hubs with a large number of connections potentially connecting several communities, like grocery stores, restaurants, religious establishments, etc.
This favors disease spread even more than randomly assigned links and can lead to outbursts of infections (superspreading events) \cite{Small2005}.
In fact, there were several super spreading events related to carnival festivities at the beginning of the first wave in Germany, for example \cite{Streeck2020}.
We believe that this, at least in part, explains why our parameter inference scheme finds larger values for the total number of social contacts $k$ than previously reported \cite{Mossong2008} for the time period before NPIs were put in place.
However, the fact that our inferred parameters for the later time periods are consistent with other reports regarding the number of contacts and the infection probability, reassures us that our assumption, that super spreading was less important after NPIs were enacted, is justified.

We inferred that lifting contact restrictions in Germany after June 6 moved the disease dynamics back into the exponential regime.
However, the effective reproduction number remained close to one, due to a reduced transmission probability of the disease, compared to the pre-lockdown time period.
This is likely a result of several phenomena: first, there might be a small seasonal effect due to the higher temperatures; second, even after contact restrictions were lifted, there was now a mask mandate at public places, and lastly, mobility was still reduced by about 20 \% compared to previous years \cite{Google} indicating a higher awareness of the virus in the population.
In October, people did spend more time indoors again, and mobility reduced to normal levels, contributing to a second fast exponential growth of cases.
The German government responded with new NPIs in November, which reduced the effective reproduction number to a value around one, again corresponding to a linear regime of new cases. 

Our analysis shows that NPIs can reduce the effective reproduction number to one by eliminating random contacts.
However, eliminating the disease does most likely require to cut almost all contacts between different cliques, for example, by working from home and not having direct contact with colleagues.
In summary, government interventions should target random contacts and encourage people to form social cliques that are disconnected from other cliques in order to efficiently prevent disease spread.

\section*{Methods}
\subsection*{Model definition}
We study an agent-based, discrete-time SEIR model on the classical Watts-Strogatz network, representing agents as network nodes.
The network is constructed by, first, connecting every node to its $k$ nearest neighbors in a ring-like topology, and, second, rewiring every link to a random node with probability $p$, see Fig.~\ref{fig:network_sketch}.
Every node $i$ has a discrete state $s_i \in \mathcal{S} = \{S, E, I, R \}$, corresponding to susceptible ($S$), exposed ($E$), infectious ($I$), and removed ($R$) states. 

Disease progression is dictated by $\Gamma$-distributed waiting times inferred from COVID-19 disease characteristics, as these have been found to describe the disease best \cite{Linton2020}.
During every discrete time step $t$, where the length of the time step corresponds to 1 day, each susceptible agent can become exposed with probability $p_I$ to every infectious agent they are connected to, 
\begin{equation}\label{eq:pinf}
  P\left(s_i(t+1) = E|s_i(t) = S\right) = 1 - (1-p_I)^{I_i},
\end{equation}
where $I_i$ is the number of infectious agents node $i$ is connected to.
Upon infection, we change the agent's state to exposed, and assign a waiting time $\tau_E \sim \Gamma(k_E, \theta_E)$ which we draw from a $\Gamma$-distribution with shape $k_E$ and scale $\theta_E$.
During every time step, the waiting times are reduced by 1 day, $\tau_E(t+1) = \tau_E(t) - 1$ if $\tau_E(t) > 0$. 
Else, the disease progresses, $s_i(t+1) = I$, and a new waiting time is assigned from another $\Gamma$-distribution $\tau_I \sim \Gamma(k_I, \theta_I)$, with shape $k_I$ and scale $\theta_I$.
Finally, when $\tau_I \leq 0$, the node is removed, $s_i(t+1) = R$.
A sketch of the SEIR dynamics can be found in Supplementary Fig.~1.
We calculate the shape and scale of the $\Gamma$-distributions from the reported mean time and variance in the respective states according to
\begin{eqnarray}
    k_{E,I} &=& \frac{\left\langle\tau_{E,I}\right\rangle^2}{\left\langle \Delta\tau_{E,I}^2 \right\rangle} , \\
    \theta_{E,I} &=& \frac{\left\langle \Delta\tau_{E,I}^2 \right\rangle}{\left\langle\tau_{E,I}\right\rangle}.
\end{eqnarray}
Note that we did not infer these parameters with our Bayesian parameter inference framework.

In our model we do not account for an inflow of infectious people by travel; we instead account for the initial surge of infections by placing randomly $n_{E}(0)$ exposed and $n_{I}(0)$ infectious agents in the population.

\subsection*{Bayesian parameter inference}
We apply approximate Bayesian computation with a sequential Monte-Carlo scheme (ABC-SMC) to infer the set of parameters $\Theta = \{ p_I, p, k, n_{E}(0), n_{I}(0) \}$ of our agent-based model.
We always keep the total number of agents fixed at $n = 3 \cdot 10^5$.
To this end, we employ the Python package pyABC \cite{Klinger2018}.
In short, the algorithm employs sequential importance sampling over generations $T=1,...,n_T$. 
In generation $T$, the algorithm draws sets of parameters $\theta_i$ from a given proposal distribution and consequently simulates data $C^{(i)}$ from the model, until $n_\mathrm{ABC}$ instances were accepted based on the comparison to observed data via a distance function $D(C^{(i)}, C_{\mathrm{obs}})$ and acceptance threshold $\varepsilon_T$, $D(C^{(i)},C_\mathrm{obs})\leq\varepsilon_T$.
As the distance function, we choose the absolute difference between new cases in the model instance $C^{(i)}(t)$ and the respective reports for Germany $C_\mathrm{obs}(t)$,
\begin{equation}
    D(C^{(i)}, C_{\mathrm{obs}}) := \sum_t \left| C^{(i)}(t) - C_{\mathrm{obs}}(t) \right|.
\end{equation}
New cases in the model are given by the daily new infections
\begin{equation}
    C^{(i)}(t) = n_I^{(i)}(t) - n_I^{(i)}(t-1) + n_R^{(i)}(t) - n_R^{(i)}(t-1).
\end{equation}
They are compared to the seven-day rolling average of new case reports in Germany $C_{\mathrm{obs}}(t)$ provided by Johns Hopkins University \cite{Dong2020} to account for the weekly fluctuations in reporting.
Acceptance of model instances depends on the acceptance threshold $\epsilon_T$ of generation $T$, which we choose as the median of the distances of the accepted instances of the previous generation
\begin{equation}
    \epsilon_{T+1} = \mathrm{median}\left( \left\{ D(C^{(i)}, C_{\mathrm{obs}}) < \epsilon_T \right\}\right).
\end{equation}
When $n_{\mathrm{ABC}}$ model instances have been accepted, the algorithm constructs new proposal distributions from the accepted instances to allow high acceptance rates while decreasing the threshold \cite{JAGIELLA2017194}.
In particular, for the continuous variables it employs a multivariate normal distribution with an adaptive covariance matrix based on the sample covariance matrix, whose scale parameter is determined by a grid search with 5-fold cross validation and refitting on the whole data set.
We compute the discrete numbers of initially exposed and infectious people $n_{E}(0), n_{I}(0)$ by rounding the continuous output of the multivariate normal distribution.
We can do this without a large error as these parameters vary smoothly over a large range.
For the parameter $k$ we employ an adaptive discrete transition that assigns probabilities to all possible parameter values directly from the frequency of the respective value in the population of accepted particles with additional random jumps (with probability 0.3) to ensure absolute continuity of the prior.
The process is repeated until the acceptance threshold is sufficiently low; we especially ensured that the threshold is considerably lower than the difference between the reported cases and the seven-day rolling average of cases.

\section*{Acknowledgments}
We thank Prof.~Jan Hasenauer for valuable discussions and input about Bayesian parameter inference. We thank the Centre for Information Services and High Performance Computing at Technische Universität Dresden for providing high-performance computing infrastructure. Diana David-Rus thanks Prof.~Manfred Wildner and the Department of Infectious Diseases Epidemiology Surveillance and Task Force at LGL-Bavaria for their support in ensuring the appropriate infrastructure.

\nolinenumbers

% Either type in your references using
% \begin{thebibliography}{}
% Text
% \end{thebibliography}
%
% or
%
% Compile your BiBTeX database using our plos2015.bst
% style file and paste the contents of your .bbl file
% here. See http://journals.plos.org/plosone/s/latex for 
\printbibliography

\end{document}

\begin{flushleft}
{\Large
\textbf{\sffamily Supplementary material for: \\
Inferring the effect of interventions on COVID-19 transmission networks}
}
\newline
% authors go here:
\\
Simon Syga\textsuperscript{1},
Diana David-Rus\textsuperscript{2},
Yannik Schälte\textsuperscript{3,4},
Michael Meyer-Hermann\textsuperscript{5,6},
Haralampos Hatzikirou\textsuperscript{1,7},
Andreas Deutsch\textsuperscript{1,*}
\\
\bigskip

\textbf{1} Center for Information Services and High Performance Computing,
Technische Universität Dresden, Nöthnitzer Straße 46, 01062 Dresden, Germany
\\
\textbf{2} Bavarian Health and Food Safety State Authority (LGL) -- Research division, 
 Veterinärstraße~2, 85764, Oberschleißheim, Germany
\\
\textbf{3} Institute of Computational Biology, Helmholtz Zentrum München –
German Research Center for Environmental Health, 85764 Neuherberg, Germany
\\
\textbf{4} Center for Mathematics, Technische Universität München, 85748 Garching, Germany
\\
\textbf{5} Department of Systems Immunology and Braunschweig Integrated Centre of Systems Biology (BRICS), Helmholtz Centre for Infection Research, Braunschweig, Germany
\\
\textbf{6} Institute for Biochemistry, Biotechnology and Bioinformatics,
Technische Universität Braunschweig, Braunschweig, Germany
\\
\textbf{7} Mathematics Department, Khalifa University, P.O. Box 127788, Abu Dhabi, United Arab Emirates
\\
\bigskip
*andreas.deutsch@tu-dresden.de
\end{flushleft}
\renewcommand\thefigure{S\arabic{figure}}
\renewcommand\thetable{S\arabic{table}}
\renewcommand\theequation{S\arabic{equation}}

\section{Bayesian parameter inference}
\subsection*{February 26 - March 15}
The early phase of the epidemic is characterized by a low number of cumulative infections.
We can therefore directly use the absolute numbers of new infections as input for our agent-based model, as we are always far away from the epidemic threshold.
We choose broad, uniform priors for all the parameters that can be found in table~\ref{tab:prior1}.
Note that the probability for random connections varies on several orders of magnitude $p \in [10^{-6}, 10^0]$ and we therefore infer this parameter on a logarithmic scale.
We use $n_{\mathrm{ABC}} = 100$ and obtain an effective sample size of $n_{\mathrm{eff}} \approx 46$.
\begin{table}[ht!]
    \centering
    \caption{\bfseries Priors of model parameters for the time period  February 26 to March 15. \label{tab:prior1}}
    \begin{tabular}{lcc}\hline
        Parameter & Variable &  Prior distribution \\\hline
        Infection probability & $p_I$ & \verb|Uniform(0.01,0.07)| \\
        Probability of random links & $\log_{10} p$ & \verb|Uniform(0,-6)|\\
        Number of links & $k/2$ & \verb|DiscreteUniform(1,16)|\\
        Initially exposed & $n_E(0)$ & \verb|Uniform(0,47)|\\
        Initially infectious & $n_I(0)$ & \verb|Uniform(0,160)|
       \\\hline 
    \end{tabular}
\end{table}

\subsection*{March 16 - June 6}
In the time period from March 16 to June 6 Germany recorded 177652 cases in total.
This means that it becomes computationally unfeasible to replicate the population directly in our model without noticing a strong effect of the removed (immune) agents. 
Therefore, we scale down the total number of infections to our system size and compare the relative number of cases per 300,000 people instead.
Assuming our hypothesis that the NPIs lead to a strongly clustered transmission network holds, we expect a large number of unconnected communities in Germany in that time period, which we represent as distinct model instances.
We use $n_{\mathrm{ABC}} = 200$ and obtain an effective sample size of $n_{\mathrm{eff}} \approx 105$. Our priors for this period can be found in table~\ref{tab:prior2}.
\begin{table}[ht!]
    \centering
    \caption{\bfseries Priors of model parameters for the time period March 16 to June 6.\label{tab:prior2}}
    \begin{tabular}{lcc}\hline
        Parameter & Variable &  Prior distribution \\\hline
        Infection probability & $p_I$ & \verb|Uniform(0.01,0.03)| \\
        Probability of random links & $\log_{10} p$ & \verb|Uniform(0,-6)|\\
        Number of links & $k/2$ & \verb|DiscreteUniform(1,11)|\\
        Initially exposed & $n_E(0)$ & \verb|Uniform(3,57)|\\
        Initially infectious & $n_I(0)$ & \verb|Uniform(38,414)|
       \\\hline 
    \end{tabular}

\end{table}
\subsection*{June 7 - September 15}
To infer the parameters of the system during this time period we first sample parameters from the posterior distribution which we obtained for the previous time period, and let the system evolve for 81 days (corresponding to the time period from March 16 to June 6).
Next, we change the infection probability $p_I$ and assign a new transmission network based on a new set of parameters $p, k$.
For these parameters, we choose the same prior distributions as for the previous time period, see Table~\ref{tab:prior3}.
We use $n_{\mathrm{ABC}} = 200$ and obtain an effective sample size of $n_{\mathrm{eff}} \approx 164$.
\begin{table}[ht!]
    \centering
    \caption{\bfseries Priors of model parameters for the time period June 7 to September 15.\label{tab:prior3}}
    \begin{tabular}{lcc}\hline
        Parameter & Variable &  Prior distribution \\\hline
        Infection probability & $p_I$ & \verb|Uniform(0.01,0.03)| \\
        Probability of random links & $\log_{10} p$ & \verb|Uniform(0,-6)|\\
        Number of links & $k/2$ & \verb|DiscreteUniform(1,11)|
       \\\hline 
    \end{tabular}
\end{table}

\section{Parameter scan}
\subsection{SEIR model}
To investigate the disease dynamics in the small-world network, we perform a parameter scan.
We vary the network parameters $p, k$ while keeping the rest of the parameters fixed at $n = 10^5, p_I = 0.02, n_E(0) = 0, n_I(0) = 10$.
Our choice for $p_I$ during the parameter scan is motivated by reports of the COVID-19 individual-level secondary attack rate (SAR) in the household of 17 \%.
Inverting Eq~3 we obtain 
\begin{equation}
    p_I = 1 - \sqrt[\tau_I]{1 - SAR} \approx 0.02.
\end{equation}
We vary the number of contacts from $k = 2, 4, \dots, 24$ and sample the probability for random contacts in eleven equally-spaced steps on the log scale from $\log_{10} p = -5, \dots, 0$.
As initial condition, 10 random agents are set to the infectious state and the system is simulated until there are no more exposed and infectious agents.
We repeat this process five times per parameter combination $(p, k)$.
As output we determine the peak of simultaneously infectious people
\begin{equation}
    n_\mathrm{peak}(p, k) := \max_t n_I(t, p, k),
\end{equation}
and the cumulative infection curves
\begin{equation}
    N(t, p, k) =  n_I(t,p,k) + n_R(t,p,k).
\end{equation}

\subsection{SIR model}
To compare our analytical prediction of the wave speed in the highly clustered network (see below), we define a simplified SIR model. 
The difference to the SEIR model is that there is no exposed state, and that the waiting times for the transition from the infectious to the removed state are drawn from an exponential distribution with mean $\ev{\tau_I} = 10 d$.
For the parameter scan, we initialze the system of $n = 10^5$ agents in a ring-like topology (no random links) with a single infectious agent and let it evolve for 365 time steps.
We repeat this process 20 times  per parameter $k$.
We then record the cumulative infection curves
\begin{equation}
    N(t, k) =  n_I(t,k) + n_R(t,k),
\end{equation}
see Fig.~5\textbf{a}. 
We calculate the linear growth rate from the cumulative infections as
\begin{equation}
    c(k) := \left\langle \frac{N(t_{\max, k}, k) - N(t_{\min}, k)}{t_{\max, k} - t_{\min}} \right\rangle,
\end{equation}
where we neglect the initial exponential growth by skipping $t_{\min} = \left\langle \tau_I \right\rangle = 10$ time steps.
We also determine the maximum time $t_{\max, k}$ until the epidemic dies out for each parameter $k$ so that after  $t_{\max, k}$ time steps, the cumulative number of infections did not increase in any realization of the system
\section{Mathematical analysis}
\subsection*{Epidemic threshold}
We can calculate an upper bound for the number of contacts $k_c$ by demanding $R_0 = 1$, i.~e. a single infectious person in a network of susceptible people will effect on average one other person.
Using Eq~3 (Main Text) we can calculate the expected number of infections as
\begin{equation}\label{eq:k_c}
    R_0 = 1 = k_c \left[1 - (1 - p_I)^{\tau_I}\right] \Longleftrightarrow k_c = \frac{1}{1 - (1 - p_I)^{\tau_I}}.
\end{equation}
For $p_I = 0.02$ and $\left\langle \tau_I \right\rangle \approx 10$ we obtain $k_c \approx 5.5$.
\subsection*{Derivation of differential-equation approximation}
In order to predict the linear growth of infections in the highly clustered small-world network, we consider a simplified variant of our original model, where we neglect the exposed state and assume that the progression times are distributed exponentially (SIR model, section 2.2 in Supplementary Information).
Then we describe the state of an agent $j$ in our model by a set of three Boolean stochastic variables $s_j (t), i_j(t), r_j(t) = 0, 1$, where $s_j + i_j + r_j = 1$ and $s_j = 1$ indicates that the agent is susceptible, $i_j = 1$ means he is infectious and if $r_j = 1$ he is removed.
We represent the event "agent $j$ becomes infectious at time $t$" by the Boolean stochastic variable $\alpha_j(t)$ with 
\begin{equation}
P\left(\alpha_j(t) = 1\right) = 1 - (1-p_I)^{I_j(t)},
\end{equation}
where $I_j(t) := \sum_{m \in \mathcal{N}_j} i_m(t)$ is the number of infectious agents in the neighborhood $\mathcal{N}_j$ of agent $j$.
Similarly, the event "agent $j$ is removed at time $t$" is given by the stochastic variable $\beta_j(t)$ with
\begin{equation}
    P\left( \beta_j(t) = 1 \right) = p_R = 1 / \tau_I.
\end{equation}
During one time step the state of all agents changes as
\begin{eqnarray}
    s_j(t+1) &=& s_j(t) - \alpha_j(t) s_j(t), \\
    i_j(t+1) &=& i_j(t) + \alpha_j(t) s_j(t) - \beta_j(t) i_j(t), \\
    r_j(t+1) &=& r_j(t) + \beta_j(t) i_j(t).
\end{eqnarray}
We want to calculate the expected value of the state variable under a mean-field approximation, i.~e. we replace the expected value of a function $f(X)$ of any random variable $X$ by the function evaluated at the expected value of the random variable, $\left\langle f(X) \right\rangle \approx f(\left\langle X \right\rangle )$.
In particular, this also means that we neglect any correlations between the state variables of neighboring nodes.
We denote the expected values of the state variables as $\sigma_j(t) := \ev{s_j(t)}, \iota_j(t) := \ev{i_j(t)}, \rho_j := \ev{r_j(t)}$.
For the expected number of infectious neighbors we obtain
\begin{equation}\label{eq:I_n}
    \ev{I_j(t)} = (1-p) \sum_{m=-k/2}^{k/2} \iota_m(t) + \frac{kp}{N-1} \sum_{m \neq j} \iota_m(t),
\end{equation}
where $N$ is the total number of nodes in the network and $p$ is the probability of a random link (see Model definition).

Next, we also introduce a small time step length $\tau > 0$ and continuous time $\hat{t} = \tau t$ along with the transition rates $\kappa_I := p_I / \tau, \kappa_R := p_R / \tau$.
In the following we only consider continuous time and drop the hat for better readability.
For $\tau \to 0$ we can approximate the expected value of $\alpha_j(t)$ by a Taylor approximation around $p_I = 0$ as
\begin{eqnarray}
    P\left(\alpha_j(t) = 1\right) = 1 - (1-\kappa_I \tau)^{I_j(t)} \approx \kappa_I \tau I_j(t).
\end{eqnarray}
We can further simplify our system by considering the regime $N\to \infty$, and introducing the spatial step length $\Delta x > 0$ so that $N \Delta x = L = \mathrm{const}$.
We can then replace the state probabilities $\sigma_j(t), \iota_j(t), \rho_j(t)$ by the probability densities $\sigma(x=j\Delta x, t), \iota(x=j\Delta x, t), \rho(x=j\Delta x, t)$.
This allows us to approximate the expected number of infectious agents in the neighborhood by expanding the terms $\iota(x+m\Delta x, t) \approx \iota(x, t) + m \Delta x \partial_x \iota(x,t) + m^2 \Delta x^2 / 2 \partial_{xx} \iota(x, t)$ to obtain
\begin{eqnarray}
    \ev{I_j(t)} &=& (1-p) \sum_{m=-k/2, m \neq 0}^{k/2} \iota_{j+m}(t) + \frac{kp}{N-1} \sum_{m \neq j} \iota_m (t) \approx \nonumber\\
    &\approx& (1-p) k \left( \iota(x,t) + \Delta x^2 \tilde{k} \partial_{xx} \iota(x,t) \right) + \frac{kp}{L} \int_0^L \iota(x,t) \, \mathrm{d}x =: \Psi (x,t),
\end{eqnarray}
where $\tilde{k} := (k/2+1) (k+1) / 12$.
Approximating all time-dependent functions by their Taylor approximation up to second order $f(t+\tau)\approx f(t) + \tau f'(t) + \tau^2 / 2 f''(t)$ and rearranging terms we obtain the following set of non-linear PDEs for the expected value of the agents' states
\begin{eqnarray}
    \partial_{tt} \sigma (x,t) + \frac{2}{\tau}\partial_t \sigma (x, t) &=& - \frac{2}{\tau} \kappa_I \sigma_j(t) \Psi (x,t), \\
    \partial_{tt} \iota (x,t) + \frac{2}{\tau}\partial_t \iota (x, t) &=& \frac{2}{\tau} \left[\kappa_I \sigma(x, t) \Psi (x,t) - \kappa_R \iota (x, t) \right], \\
    \partial_{tt} \rho (x,t) + \frac{2}{\tau} \partial_t \rho (x, t)) &=& \frac{2}{\tau} \kappa_R \iota (x, t).
\end{eqnarray}
We introduce the constant $D_k := \frac{\Delta x^2 \tilde{k}}{\tau}$ and can now identify two separate time scales: A fast time scale, with terms $\propto \frac{1}{\tau}$ which characterizes the disease progression, and a slow timescale with terms $\propto D_k$ that describes the wave of infections in the network.
Finally, with $I(t) := 1/L \int_0^L \iota(x,t) \, \mathrm{d}x$ as the total number of infectious people, we obtain the following set of nonlocal PDEs
\begin{eqnarray}\label{eq:fullPDE}
    \partial_{tt} \sigma (x,t) &=&  -\frac{2}{\tau} \left\{ \partial_t \sigma (x, t) + \kappa_I k \sigma (x,t) \left[ (1-p) \left( \iota(x,t)  \right) + p I(t) \right]\right\} -\kappa_I k \sigma (x,t)D_k \partial_{xx} \iota(x,t), \\ \label{eq:iotapde}
    \partial_{tt} \iota (x,t) &=& \frac{2}{\tau} \left\{ -\partial_t \iota (x, t) +\kappa_I k \sigma (x,t) \left[ (1-p) \left( \iota(x,t)  \right) + p I(t) \right] - \kappa_R \iota(x,t) \right\} + \nonumber \\
    && + \kappa_I k \sigma (x,t) D_k \partial_{xx} \iota(x,t), \\
     \partial_{tt} \rho (x,t) &=& \frac{2}{\tau} \left\{- \partial_t \rho (x, t)) + \kappa_R \iota(x,t) \right\}.
\end{eqnarray}
In the regime $\tau \to 0$ we expect the the system to quickly reach a local steady state, so that the fast time scale terms vanish.
We can use this to obtain an approximate value for the wave speed of infections.
\subsection*{Calculation of the infection wave speed in the highly clustered network}
To calculate the wave speed of infections we consider the regime $p=0$, so that we can neglect the nonlocal coupling by $I(t)$ in Eq~\ref{eq:fullPDE}.
As described above, if we let $\lim \tau, \Delta x \to 0$ with finite $D_k$, we expect the system to always be in a local steady state, corresponding to the fast time scale terms being 0.
Then, we arrive at the following equation for the density of infected people
\begin{equation}\label{eq:wavelike}
\partial_{tt} \iota(x,t) = \kappa_I k D_k \sigma(x,t) \partial_{xx} \iota(x,t).    
\end{equation}
If we consider a point far away from the wave front, we have $\sigma(x,t) \approx 1$ (everyone is susceptible), and Eq~\ref{eq:wavelike} reduces to the wave equation
\begin{equation}
    \partial_{tt} \iota(x,t) = \kappa_I k D_k \partial_{xx} \iota(x,t)    
\end{equation}
with the wave speed $c = \sqrt{\kappa_I k D_k} \propto k\sqrt{k}$.
\section*{Supplementary Figures}
\begin{figure}
    \centering
    \includegraphics[width=\textwidth]{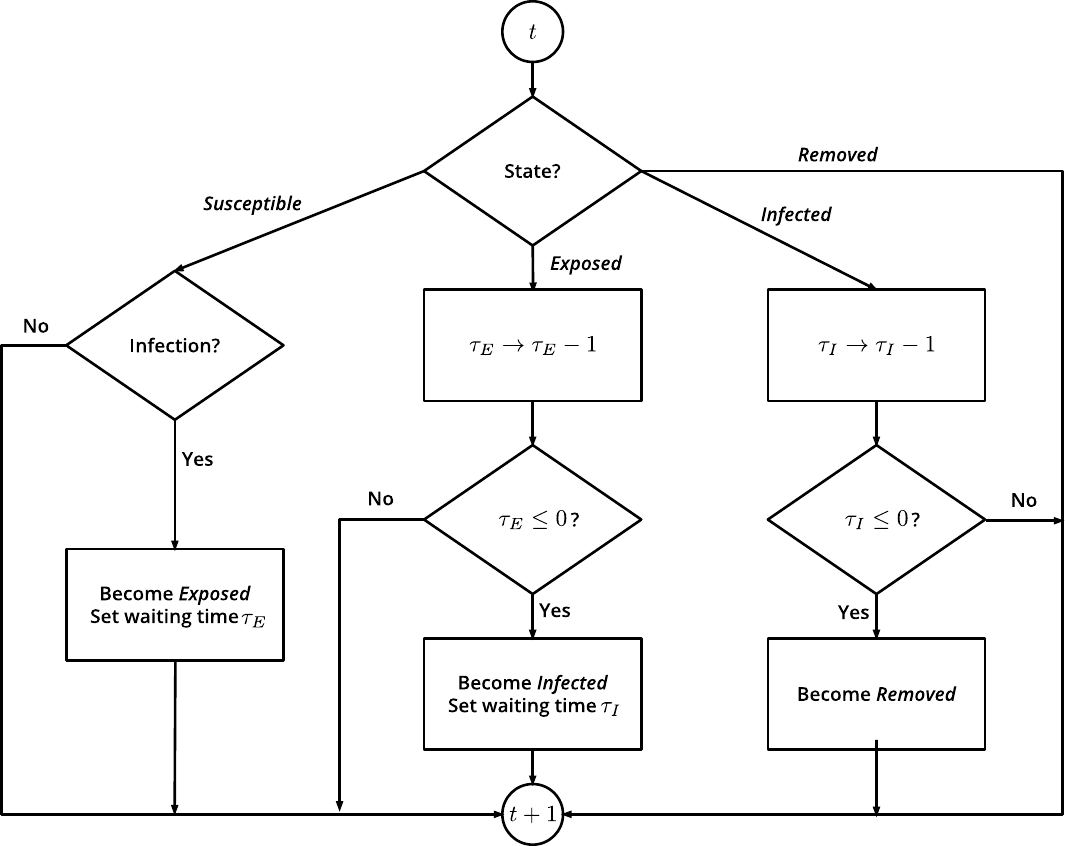}
    \caption{\textbf{Dynamics in the agent-based SEIR model.} Susceptible agents can become exposed, if they are linked to infectious agents. Exposed agents become infectious after the waiting time $\tau_E$, and infectious agents are removed after $\tau_I$. All nodes are updated simultaneously at every time step $t$.}
    \label{fig:S2_Fig}
\end{figure}
% \printbibliography